\documentclass[prb, 10pt,twocolumn]{revtex4-1}
\usepackage{enumerate} 	
\usepackage{amsmath}	
\usepackage[T1]{fontenc}
\usepackage{graphicx}	
\usepackage{amssymb}

\newcommand{\ket}[1]{|#1\rangle} 	
\newcommand{\bra}[1]{\langle#1|} 	

\newcommand{\bracket}[2]{\langle#1|#2\rangle}

\begin{document}

\title{Green Functions of Graphene: An Analytic Approach}

\author{James A. Lawlor}
\affiliation{School of Physics, Trinity College Dublin, Dublin 2, Ireland}

\author{Mauro S. Ferreira}
\affiliation{CRANN, Trinity College Dublin, Dublin 2, Ireland}
\affiliation{School of Physics, Trinity College Dublin, Dublin 2, Ireland}

\begin{abstract}
 In this article we derive the lattice Green Functions (GFs) of graphene using a Tight Binding Hamiltonian incorporating both first and second nearest neighbour
 hoppings and allowing for a non-orthogonal electron wavefunction overlap. 
 It is shown how the resulting GFs can be simplified from a double to a single integral form to aid computation, and that 
 when considering off-diagonal GFs in the high symmetry directions of the lattice this single integral can be approximated very accurately by an algebraic expression.
 By comparing our results to the conventional first nearest neighbour model commonly found in literature, it is apparent that
 the extended model leads to a sizeable change in the electronic structure away from the linear regime. 
 As such, this article serves as a blueprint for researchers who wish to examine quantities where these considerations are important.
\end{abstract}

\maketitle

\section{Introduction}

 Green Functions (GFs) are useful tools for describing the electronic structure of materials and various other quantities related to the electronic density of a material,
 such as the local density of states, inter-impurity interactions and scattering processes.
 A common strategy is to find a suitable Hamiltonian for the material and then obtain the system GFs computationally. 
 Although the result of such an approach is accurate it can be computationally expensive and misses the finer mathematical details often masked by numerical intricacies. 
 Hence Green Functions can become more useful when they can be expressed in a simple mathematical form. 
 There has been extensive work done on graphene-based materials and the GFs of these systems throughout the years.
 In particular, analytic expressions for the GFs of graphene have been derived and used to explain such phenomena 
 as magnetic coupling between impurities \cite{sherafati, stephen_rkky, Klinovaja} and Friedel Oscillations\cite{me_friedel, bacsi}. 
 In those references, the GFs were obtained for a single-orbital tight binding model based on orthogonal states.

\begin{figure}[!ht]
\centering
 \includegraphics[scale=0.4]{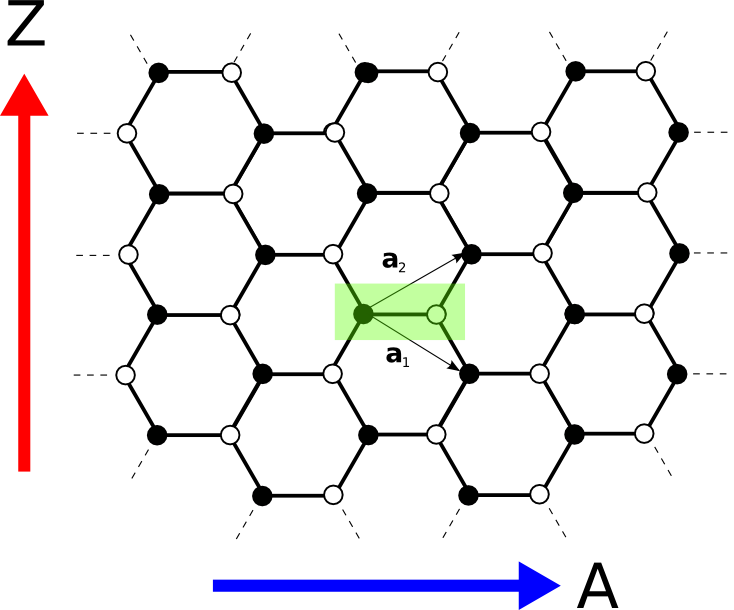}
 \caption{Schematic of a small part of the graphene lattice showing the primitive lattice vectors 
$\mathbf{a}_1 = \left\{\frac{3a}{2}\mathbf{\hat{x}},\frac{\sqrt{3}a}{2}\mathbf{\hat{y}}\right\}$ and $\mathbf{a}_2 = \left\{\frac{3a}{2}\mathbf{\hat{x}},-\frac{\sqrt{3}a}{2}\mathbf{\hat{y}}\right\} $ and
the two atom unit cell enclosed by the green transparent box. 
Using these vectors the location of any unit cell in the lattice is defined as $\mathbf{r} = m \mathbf{a}_1 + n \mathbf{a}_2$
where $m , n \in \mathbb{Z}$.
 The armchair and zigzag directions are indicated by A and Z respectively and will be used to specify directions for the Green Functions used later in this work.}
\label{fig:unitcell}
\end{figure}

In this paper we show how the single-particle lattice Green Functions can be found for graphene 
using a second nearest neighbour non-orthogonal Tight Binding model. 
 It is common in the literature to find first nearest neighbour approximations, as well as the assumption of an orthogonal basis for the electron wavefunctions \cite{arrieta,Bena2,wehling_tbm,sherafati,kogan, stephen_rkky, Rodrigues,lherbier}. 
 Work by Reich et al. \cite{reich} showed that this approximation is only really valid in the linear regime and that an improvement can be made by including further interactions.
 Our motivation is to improve the previous GF results by accounting for the extended electron hoppings and wavefunction overlaps.
 Extending the model in this way has already shown to be necessary for the electronic structure of nanoribbons \cite{Wu_gnr}.
 Furthermore we will show that previous methods for approximating the off-diagonal GFs via the Stationary Phase Approximation \cite{stephen_spa} are applicable to this extended case also.
 These approximations of the off-diagonal GFs improve in accuracy with increasing separations,
 and as such are perfectly suited for in-depth analysis of long-range phenomena in graphene such as the RKKY interaction \cite{stephen_rkky,sherafati,kogan} and Friedel Oscillations \cite{bacsi,me_friedel},
 for energies outside the linear spectrum.
  
The paper is organised as follows.
Firstly there is an introduction to general GF methods, followed by 
a derivation of the associated lattice GFs for graphene in integral form.
 Finally, we show how these integrals can be approximated to a high degree of accuracy in the high symmetry directions of the lattice,
 allowing for a fully analytic expression of the associated GFs. 

\section{Methodology}
\subsection{Tight Binding Hamiltonian}

To apply the techniques of Green Functions to graphene, we must first derive a Hamiltonian to describe the system.
 This can be done by applying a second nearest-neighbour tight binding model, which assumes that electrons can hop from one atomic site to its first and second nearest neighbours in the lattice.
 Saito et al. \cite{Saito} derived the dispersion relation using first nearest neighbour hopping and overlap. 
 In this section we will show, in detail, how a similar approach can be used to obtain the dispersion relation with the inclusion of the second nearest neighbour interactions,
 and further we identify the eigenvectors of this system.
The graphene lattice is composed of two triangular and inter-penetrating sublattices 
which we will refer to as Black ($\bullet$) and White ($\circ$) and we choose a 2-atom unit cell as shown in Fig. \ref{fig:unitcell}.
 Our assumption means that each atom has three first-nearest neighbours belonging to the opposite sublattice, and six second-nearest neighbours belonging to the same sublattice.
 The wavefunction overlap is assumed to exist between first-nearest neighbours only.

This system is described by the real-space Hamiltonian

\begin{widetext}
\[ H = \epsilon_0 \left( \sum_\mathbf{r}  \ket{\mathbf{r},\bullet} \bra{\mathbf{r},\bullet} +  \ket{\mathbf{r},\circ} \bra{\mathbf{r},\circ} \right) \]
\[ + t \left( \sum_\mathbf{r}  \ket{\mathbf{r},\bullet} ( \bra{\mathbf{r},\circ} + \bra{\mathbf{r-a_2} , \circ} + \bra{\mathbf{r-a_1}, \circ} )
 + \ket{\mathbf{r},\circ} ( \bra{\mathbf{r}, \bullet} +  \bra{\mathbf{r+a_2} , \bullet} + \bra{\mathbf{r+a_1},\bullet} \right)\]
\begin{equation}
  + t' \Bigg( \sum_\mathbf{r} \ket{\mathbf{r},\bullet} (\bra{\mathbf{r-a_1},\bullet} +  \bra{\mathbf{r-a_2},\bullet} + \bra{\mathbf{r+a_1-a_2},\bullet} 
 +  \bra{\mathbf{r+a_1},\bullet} +  \bra{\mathbf{r+a_2},\bullet} +  \bra{\mathbf{r+a_2-a_1},\bullet}) 
 \label{eq:hamiltonian}
 \end{equation}
\[ +  \ket{\mathbf{r},\circ} ( \bra{\mathbf{r+a_1},\circ} +  \bra{\mathbf{r+a_2},\circ}  +   \bra{\mathbf{r+a_2-a_1},\circ} \\
  +  \bra{\mathbf{r-a_1},\circ}  +  \bra{\mathbf{r-a_2},\circ}  +  \bra{\mathbf{r+a_1-a_2},\circ} ) \Bigg)  \\ \]
\end{widetext}

where the vector $\mathbf{r}$ is defined as per Fig. \ref{fig:unitcell} and is summed over the entire system to infinity. 
The parameters $t$ and $t'$ are negative energies denoting the first and second nearest neighbour hopping integrals respectively,
and $\epsilon_0$ corresponds to the on-site energy of each carbon atom.

 Accurate values of these parameters can be found through Density Functional Theory, and many examples exist in the literature \cite{reich, kundu}. 
 Numerical calculations throughout this paper will use units of the first nearest neighbour hopping $t = -1$,
 and using the parameterisation of S. Reich\cite{reich} gives $t' = -0.037$ and $\epsilon_0 = 0.111$. 
The real-space Hamiltonian can be diagonalised using a Fourier Transform from real- to reciprocal-space

\[ \ket{\mathbf{k},A} = \frac{1}{\sqrt{N}} \sum_r e^{i \mathbf{k.r}} \ket{\mathbf{r},A}, \]

with the inverse transform given by 

\[ \ket{\mathbf{r},A} = \frac{1}{\sqrt{N}} \sum_k e^{- i \mathbf{k.r}} \ket{\mathbf{k},A}. \]

Here, $N$ is the number of atomic sites. Such a transform results in the equivalent diagonalised Hamiltonian in k-space given by

\begin{equation}
 \hat{H}(\mathbf{k}) = \left( \begin{array}{cc}
\epsilon_0 + t' g(\mathbf{k})  & t f(\mathbf{k}) \\
t f^* (\mathbf{k}) & \epsilon_0 +  t' g(\mathbf{k}) \end{array} \right)
\label{eq:diagonalisedH}
\end{equation}

where $f(\mathbf{k}) = 1 + e^{i \mathbf{k.a_1}} + e^{i \mathbf{k.a_2}}$ and $g(\mathbf{k}) = 2 \left( \cos( \mathbf{k.a_1}) + \cos( \mathbf{k.a_2}) + \cos(\mathbf{k.a_1} + \mathbf{k.a_2}) \right)$.

\subsection{Eigenvalues and Eigenvectors of the Diagonalised Hamiltonian}

The eigenvalues of the Hamiltonian, corresponding to its spectrum, are found through applying the generalised secular equation $\det (\hat{H} - \epsilon \hat{S}) = 0$.
The matrix $\hat{S}$ is the wavefunction overlap matrix which can be written explicitly in diagonalised form as

\[ \hat{S}(\mathbf{k}) = \left( \begin{array}{cc}
1  & s f(\mathbf{k}) \\
s f^* (\mathbf{k}) & 1 \end{array} \right).\]

 Here, $s$ is a dimensionless parameter which quantifies the wavefunction overlap of neighbouring $p_z$ orbitals above each carbon site in the graphene lattice. 
 For the purposes of later calculations we will adopt the value of $s = 0.1$ from S. Reich \cite{reich}.
This non-orthogonality of the wavefunctions is commonly expressed mathematically as 
\[ \bracket{\phi_a}{\phi_b} = s \]
where $a$ and $b$ are neighbouring lattice sites.
The spectral solutions to the secular equation are

\[\epsilon_{\pm} = \frac{\epsilon_0 + t' g \pm t  \left|f\right|}{1 \pm s  \left|f\right| } .\]

It is straightforward to identify the eigenvectors of the system via Schrodinger's Equation, $\hat{H} \ket{\Psi} = \epsilon \hat{S} \ket{\Psi}$.
Assuming $ \ket{ \Psi} = \left( \begin{array}{c} A_1 \\ A_2 \end{array} \right)$ gives the matrix equation

\[ \left( \begin{array}{cc}
\epsilon_0 + t' g  & t f \\
t f^*  & \epsilon_0 + t' g \end{array} \right)  \left( \begin{array}{c} A_1 \\ A_2  \end{array} \right) = \epsilon_\pm  \left( \begin{array}{cc}
1  & s f \\
s f^*  & 1 \end{array} \right) \left( \begin{array}{c} A_1 \\ A_2  \end{array} \right). \]

Solving this and normalising through the requirement $\bracket{\Psi_\pm}{\Psi_\pm} = 1$ we find

\[ \ket{ \Psi_\pm} = \frac{1}{\sqrt{2}} \left( \begin{array}{c} 1 \\ \pm e^{\- i \phi}  \end{array} \right) \]
so
\[ \ket{\mathbf{k} , \pm} = \frac{1}{\sqrt{2N}} \sum_r e^{- i  \mathbf{k.r}} ( \ket{\mathbf{r},\bullet} \pm e^{-i \phi} \ket{\mathbf{r}, \circ} ) . \]

\section{Derivation of Lattice Green's Functions}
\label{sec:double_integral}

In a general sense Green Functions represent the impulse response of a system to a singular input.
 Specifically, in quantum physics they are used to describe the propagation of an electron through the system.
They act as an inverse to the Hamiltonian and once known contain a great deal of information about the system, 
the primary examples being the density of states, scattering processes and inter-impurity interactions.
 Impurities can be introduced to a pristine system via a perturbation approach, this is well understood and beyond the scope of this paper \cite{me_friedel}.
For a system with Hamiltonian $\mathcal{H}$, the Green Function operator $\mathcal{G}$ corresponding to the Schrodinger equation 
and describing electronic propagation is defined as \cite{economou}
\[ \hat{G} = \lim_{\eta \to 0} \sum_\mathbf{k} \sum_\pm \frac{ \ket{\mathbf{k}_\pm} \bra{\mathbf{k}_\pm}}{(E+i\eta) - \epsilon_\pm (\mathbf{k})}. \]
 For brevity the infinitesimal imaginary part $i \eta$ will be absorbed into the energy $E$ from this point onwards. 
 It is useful to change the general reciprocal space vector $\mathbf{k} = k_x \mathbf{k_x} + k_y \mathbf{k_y}$ using the dimensionless forms
 \[ k_A = \frac{a k_x}{2} \]
 \[ k_Z = \frac{\sqrt{3} a k_y}{2} \]
 in order to simplify the resulting mathematics below.
 Piecing together all the elements of the calculation so far and applying them to the second nearest neighbour tight binding Hamiltonian in Eq.\ref{eq:diagonalisedH} 
 it is possible to write the system Green's Function as
 \begin{widetext}
\begin{equation}
 g_{jl} = \frac{1}{2 \pi^2} \int_{-\pi/2}^{\pi/2} dk_z \int_{-\pi}^\pi dk_A  \sum_\pm \frac{ \gamma_\pm (1 \pm s\left|f\right|)^2 e^{i k_A (m + n) + i k_Z (m - n)} }{E(1 \pm s \left|f\right|) - \epsilon_0 - t' g \mp t \left|f\right|} 
 \label{eq:gfdouble}
 \end{equation}
 \end{widetext}
 using $\mathbf{r_l - r_j} = m \mathbf{a_1} + n \mathbf{a_2}$ as per Fig. \ref{fig:unitcell} and the $\gamma_\pm$ functions are defined as
\[ \gamma_\pm (E) = \begin{cases}  	
	1 \pm s \left|f\right| 		 &\mbox{for $g_{j,l}^{\bullet, \bullet}$ and $g_{j,l}^{\circ, \circ}$} \\ 
	s f^* \pm e^{i \phi} 	 & \mbox{for $g_{j,l}^{\bullet, \circ} $} \\
	s f \pm e^{- i \phi} 	 & \mbox{for  $g_{j,l}^{\circ, \bullet}$} \end{cases}. \]

To facilitate calculation, the integration over the hexagonal First Brillouin Zone in k-space has been deformed to an equivalent rectangular area,
which is possible through symmetry considerations \cite{stephen_spa}.
Although $g_{jl}$ in its current form completely describes the second nearest neighbour system with overlap, its calculation is computationally expensive due to the 
double integral.
In the next section it will be shown how one of the integrals in Eq.\ref{eq:gfdouble} can be solved analytically through contour integration, providing an equivalent solution
with an improvement of 1-2 orders of magnitude in computation time. 
 We further demonstrate that the resulting single integral Green's Function can be approximated in certain cases to yield a fully analytic 
 expression for the system Green's Functions. 

\begin{figure}[!ht]
\centering
 \includegraphics[scale=0.4]{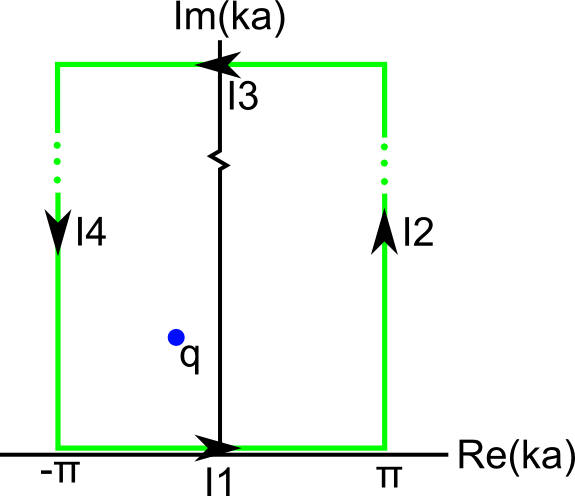}
 \caption{Integration contour (green) chosen in the complex $k_A$ plane with an example of a pole in the integrand ($q$) indicated by the blue circle. 
 The integral $I_3$ is taken as $ Im (k_A) \to \infty$.
 It follows from residue theorem \cite{saff} that the sum of the integrals $I_1 + I_2 + I_3 + I_4$ is equal to $2 \pi i$ multiplied by the sum of residues within the contour. }
\label{fig:integration_contour}
\end{figure}

\subsection{Single Integral expression for the Lattice Green's Functions}
\label{sec:single_integral}

 The methods of complex analysis , in particular contour integration \cite{saff}, can be used to solve either one of the two integrals in Eq. \ref{eq:gfdouble} 
 and in this section the method to evaluate the integral over $k_A$ will be demonstrated.
 The resulting expression is more mathematically complicated than the integral version,
 but quicker to calculate computationally, a distinct advantage over ab-initio based approaches.

 Identifying the integral to be evaluated as 
 \begin{equation}    
I_1 = \frac{1}{2 \pi^2} \int_{-\pi/2}^{\pi/2} dk_A \sum_\pm \frac{ \gamma_\pm (1 \pm s\left|f\right|)^2 e^{i k_A (m + n) + i k_Z (m - n)} }{E(1 \pm s \left|f\right|) - \epsilon_0 - t' g \mp t \left|f\right|}
\label{eq:I1}
 \end{equation}
 we can form a contour in the complex $k_A$ plane as shown in Fig.\ref{fig:integration_contour}.

 In general a function of the form $\frac{u(x)}{v(x)}$ has poles (i.e. the function diverges to infinity) for values $x_0$ where the denominator $v(x_0) \to 0$,
 so-called simple poles.
 Via the residue theorem we can relate the sum of the integrals along the contour in Fig.\ref{fig:integration_contour} to the 
 value of the integral over $k_A$ that we are interested in finding.
 The residues of the function $\frac{u(x)}{v(x)}$ with simple poles at the points $x_0$ are defined as $Res \frac{u(x_0)}{v'(x_0)}$ where $v'$ is the first differential of $v$ with respect to $x$.
Thus for the integral in Eq.\ref{eq:I1} a simple pole $q$ occurs whenever the condition 
\[ E(1 \pm s \left|f\right|) -\epsilon_0 -  t' g \mp t \left|f\right| = 0 \]
is satisfied for values of $k_A$. 
 Whether the plus or minus sign is taken, they yield the same solutions for $q$,
\begin{widetext}
\[ \cos q = \frac{1}{8 t'^2 \cos{k_z}} \Bigg( 
 (E^2 s^2 - 2Est+t^2+2Et'- 2 \epsilon_0 t' +4t'^2) - 8t'^2\cos^2{k_z}  \pm (t-Es) \sqrt{E^2s^2-2Est+t^2+4Et'-4 \epsilon_0 t' +12t'^2} \Bigg) . \]
\end{widetext}
 However when the inverse cosine is taken the choice of sign is made such that $Im(q) > 0$, ensuring the pole exists in the contour.
It should be emphasised that the sign results from the square root and is independent of the $\pm$ choice in the sum of Eq.\ref{eq:I1}.

Concerning the sum of the integrals $I_1 + I_2 + I_3 + I_4$ shown in Fig.\ref{fig:integration_contour} it can be shown 
that $I_2 = -I_4$, which can be seen from the periodicity of the integrand. 
When $m = n \neq 0$ the exponential $ e^{i k_A (m + n)}$ vanishes along the top of the contour indicated by $I_3$.
However, care must be taken in the case $m = n = 0$ where the integrand $I_3$ no longer vanishes
 and it can be shown through functional analysis that it is finite, going as $\sim -E s$ and therefore arises as a result of the wavefunction overlap. 
 Although an exact analytic solution of the integral is not obtainable, it can either be calculated easily through computational or via functional form for such a linear solution.
 Due to the symmetry of the lattice such special care is only needed for the diagonal Green's Function $g_{00}^{\bullet \bullet}$ when $s \neq 0$, as
 $g_{00}^{\bullet \circ}$ or $g_{00}^{\circ \bullet}$ can be found using non-zero values of $m$ and $n$ thus $I_3$ vanishes.
 Hence the integral $I_1$ can otherwise be shown to be equal to
 \[ I_1 = 2 \pi i \sum_{\pm} \sum_{q} \frac{\gamma_\pm \left|f\right| (1 \pm s \left|f\right|)^2  e^{i q (m + n) + i k_Z (m - n)}}{2 \cos k_z \sin q (2 t \left|f\right| \pm (t - Es))}  \]
 and so
 \begin{equation}
  g_{jl} = A + \frac{i}{2 \pi} \int_{-\pi/2}^{\pi/2} d k_Z \sum_q \sum_\pm  \frac{\gamma_\pm \left|f\right| (1 \pm s \left|f\right|)^2  e^{i q (m + n) + i k_Z (m - n)}}{\cos k_z \sin q (2 t \left|f\right| \pm (t - Es))}
  \label{eq:semi_analytic_gf}
  \end{equation}
 where 
 \begin{equation}
 A = - \frac{1}{2 \pi^2} \int_{-\pi}^{\pi} d x \int^{-\pi/2}_{\pi/2} \sum_\pm \frac{(1 \pm s  f_A)^3 d k_Z }{E(1 \pm sf_A) - \epsilon_0 - t'(f_A^2 -3) \mp tf_A} 
 \label{eq:Acorrection}
\end{equation}
for the diagonal $g_{00}^{\bullet \bullet}$ case only, and $A = 0$ otherwise.
Additionally,  
  \[ f_A = \lim_{y \to \infty}  \sqrt{1 + 4 \cos{(x+iy)} \cos{k_Z} + 4 \cos^2{k_Z}} .\]
 
\begin{figure}[!ht]
\centering
 \includegraphics[scale=0.1]{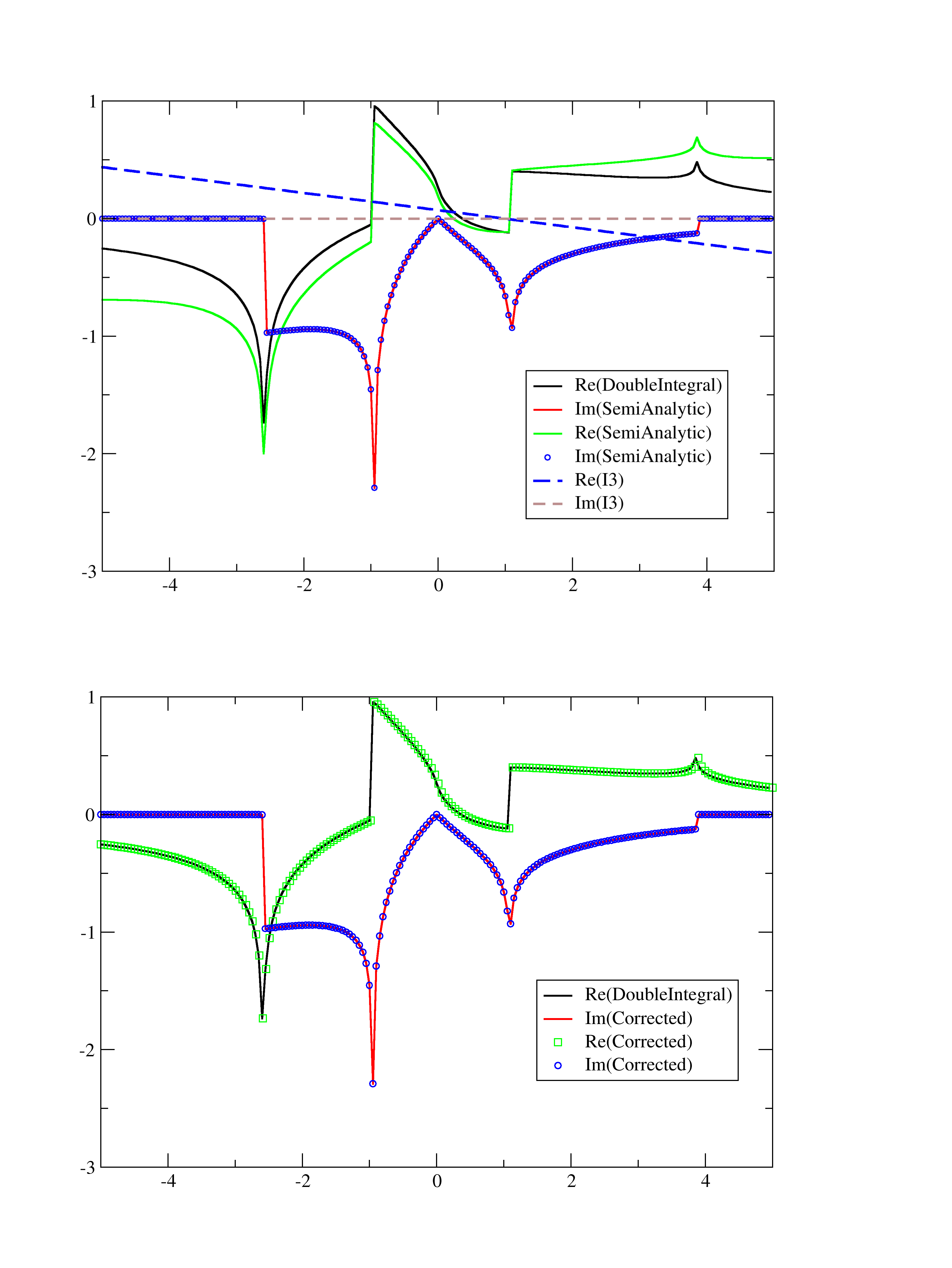}
 \caption{Comparisons of the Green's Function $g_{00}^{\bullet \bullet}$ calculated through the Double Integral and Single Integral, both without (top) and with (bottom)
 $I_3$ correction. Also shown in the left diagram is the value of $A$ from Eq.\ref{eq:Acorrection}. }
\label{fig:I3contribution}
\end{figure}

The necessity of including the extra integral $A$ for the diagonal Green's Function is demonstrated clearly in Fig.\ref{fig:I3contribution}.
 This figure also demonstrates that the single integral version of $g_{jl}$ of Eq. \ref{eq:semi_analytic_gf} agrees exactly with the double integral version of Eq. \ref{eq:gfdouble}, 
 with a reduced computation time of around 2 orders of magnitude.

\begin{figure}[!ht]
\centering
 \includegraphics[scale=0.3]{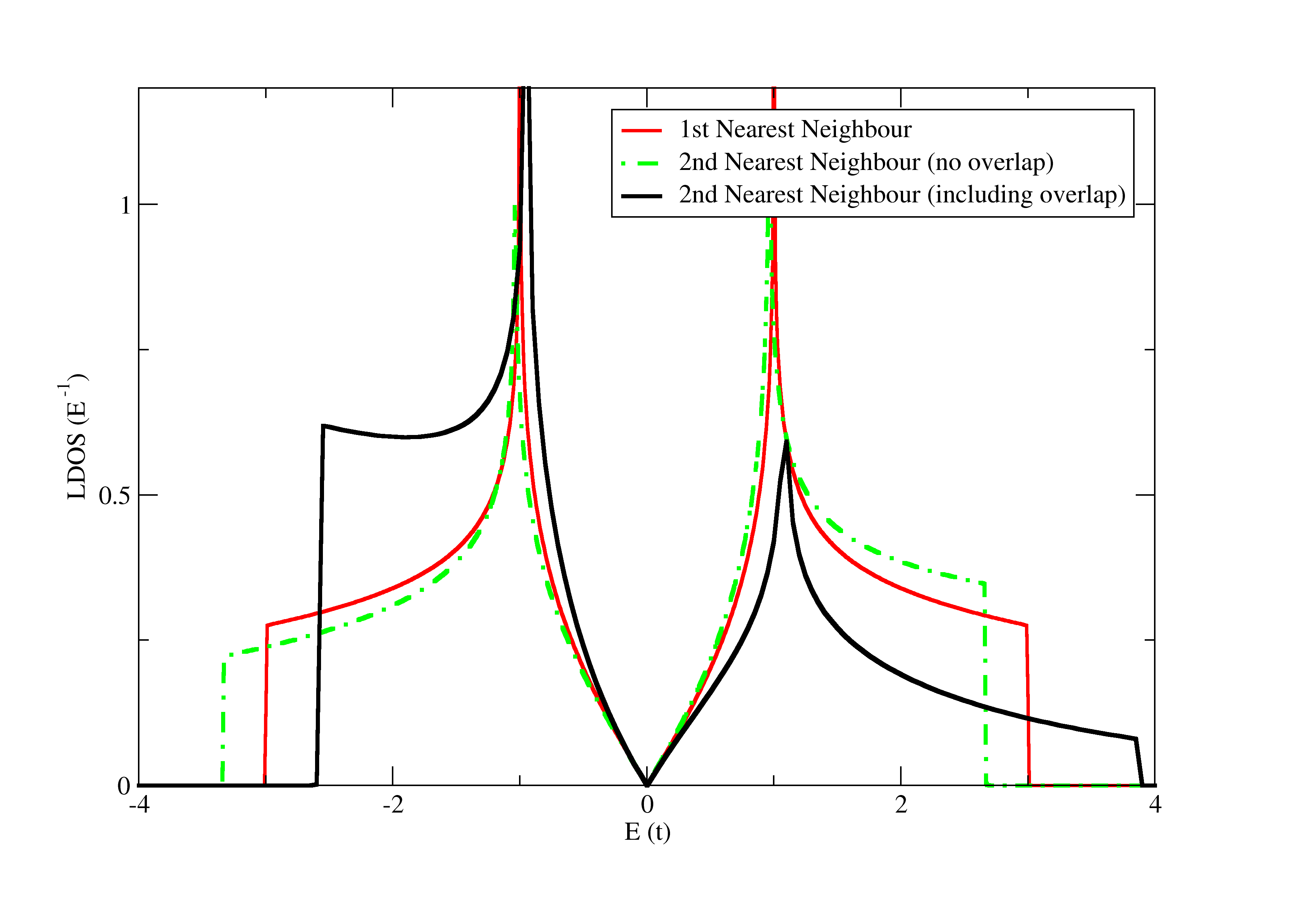}
 \caption{ Local Density of States (LDOS) spectrum comparison of the 1st Nearest Neighbour and Second Nearest Neighbour models, the latter shown for the cases $s=0$ and $s=0.1$.}
\label{fig:ldosplots}
\end{figure}

It is useful to compare the calculated Local Density of States (LDOS) between the different tight binding models, first nearest neighbour, second nearest neighbour without overlap and with overlap.
 A plot of the LDOS spectrum of the three models is shown in Fig.\ref{fig:ldosplots}. 
 Although the behaviour around the Dirac Point is similar between all models,
 it can be seen that inclusion of second nearest neighbour interactions in the Hamiltonian destroys the electron-hole symmetry in the electronic structure, 
 furthermore there is a drastic change in the spectrum when wavefunction overlap is accounted for, as shown by the black solid line.

\section{ Application to Carbon Nanotubes }

It is a straightforward exercise to use the Green's Function for graphene that was derived in Eq. \ref{eq:gfdouble} or Eq. \ref{eq:semi_analytic_gf} to
obtain the Green's Function for a Carbon Nanotube (CNT) with the second nearest neighbour and wavefunction overlap accounted for.
 This can be done through the quantization of the momentum in the circumferential direction of the desired tube, as
 only electron wavefunctions with such momenta are physically allowed.
 We will derive the Green's Function of a zig-zag CNT, so-called because the circumferential direction lies in the zig-zag direction as indicated in Fig. \ref{fig:unitcell}, but
 indeed this methodology can be applied to other chiralities.
 Zig-zag CNTs are particularly interesting because their electronic characteristics are uniquely determined by their width $N_C$, where $N_C$ is an integer 
 indicating how many unit cells are in the circumferential direction.
 It is well known that such nanotubes are metallic for widths $N_C = 0 \bmod{3}$, and semi-conducting otherwise \cite{cnts}. 
 In the final subsection the band gap obtained using this method will be compared to the first nearest neighbour approach, a recursive Green's Function approach and recent DFT results from
 the literature \cite{b3lyp}.
 
 \subsection{ Derivation of the Green's Function of a Zig-Zag Carbon Nanotube }
 
 Beginning with Eq. \ref{eq:gfdouble}, the momentum in the $k_Z$ direction for a zig-zag CNT must be quantized such that $k_Z = \frac{\pi j}{N_C}$, where $j$ is an integer running from
 $0$ to $N_C - 1$.
 Consequently the integral over the Brillouin Zone must be adapted from $ \frac{1}{\pi} \int d k_Z \to \frac{1}{N_C} \sum_j$ to account for this.
 The resulting single integral expression can further be solved using the methods of Section \ref{sec:single_integral}, yielding a fully analytic and exact
 expression for the CNT Green's Function
 \[ g_{jl} = \frac{i}{2 N_C} \sum_{k_Z} \sum_q \sum_\pm  \frac{\gamma_\pm \left|f\right| (1 \pm s \left|f\right|)^2  e^{i q (m + n) + i k_Z (m - n)}}{\cos k_z \sin q (2 t \left|f\right| \pm (t - Es))}. \]
 The poles here are similar to those of Eq. \ref{eq:semi_analytic_gf} but with the quantized $k_Z$ momentum.
 
 \subsection{ Comparison of the resulting band gap to other methods }

\begin{table*}
  \centering
\scalebox{1.4}{
  \begin{tabular}{|c||c|c|c|c|}
  \hline
$N_C$ & 1st N.N. & 2nd N.N. w/Overlap & DFT & 1st N.N. (Recursive) \\
\hline
7  & 1.358 eV (146.5\%) & 1.377 eV (148.6\%) & 0.927 eV (100\%) & 1.2609 eV (136.0\%) \\
11 & 0.929 eV (81.6\%)  & 0.942 eV (82.7\%)  & 1.139 eV (100\%) & 0.8613 eV (75.6\%)  \\
13 & 0.748 eV (86.5\%)  & 0.759 eV (87.7\%)  & 0.865 eV (100\%) & 0.6939 eV (80.2\%)  \\
17 & 0.596  eV (81.2\%)  & 0.604 eV (82.4\%)  & 0.734 eV (100\%) & 0.5535 eV (75.4\%)   \\
19 & 0.516 eV (83.7\%)  & 0.523 eV (84.9\%)   & 0.617 eV (100\%) & 0.4779 eV (77.5\%) \\
\hline \end{tabular}}
  \caption{Comparison of the calculated band gap for various semi-conducting zig-zag CNT using different methods. 
  Each value has in parentheses its percentage difference from the DFT calculated values \cite{b3lyp} for that particular CNT.
  An important caveat of the band gaps obtained from the tight-binding method is their strong dependence on the parameterisation of the system. 
  The values in this table were calculated using $t = 2.7$eV, $t' = 0.1$eV and $s = 0.1$.
  }
  \label{tab:1}
\end{table*}

\begin{figure}[!ht]
\centering
 \includegraphics[scale=0.35]{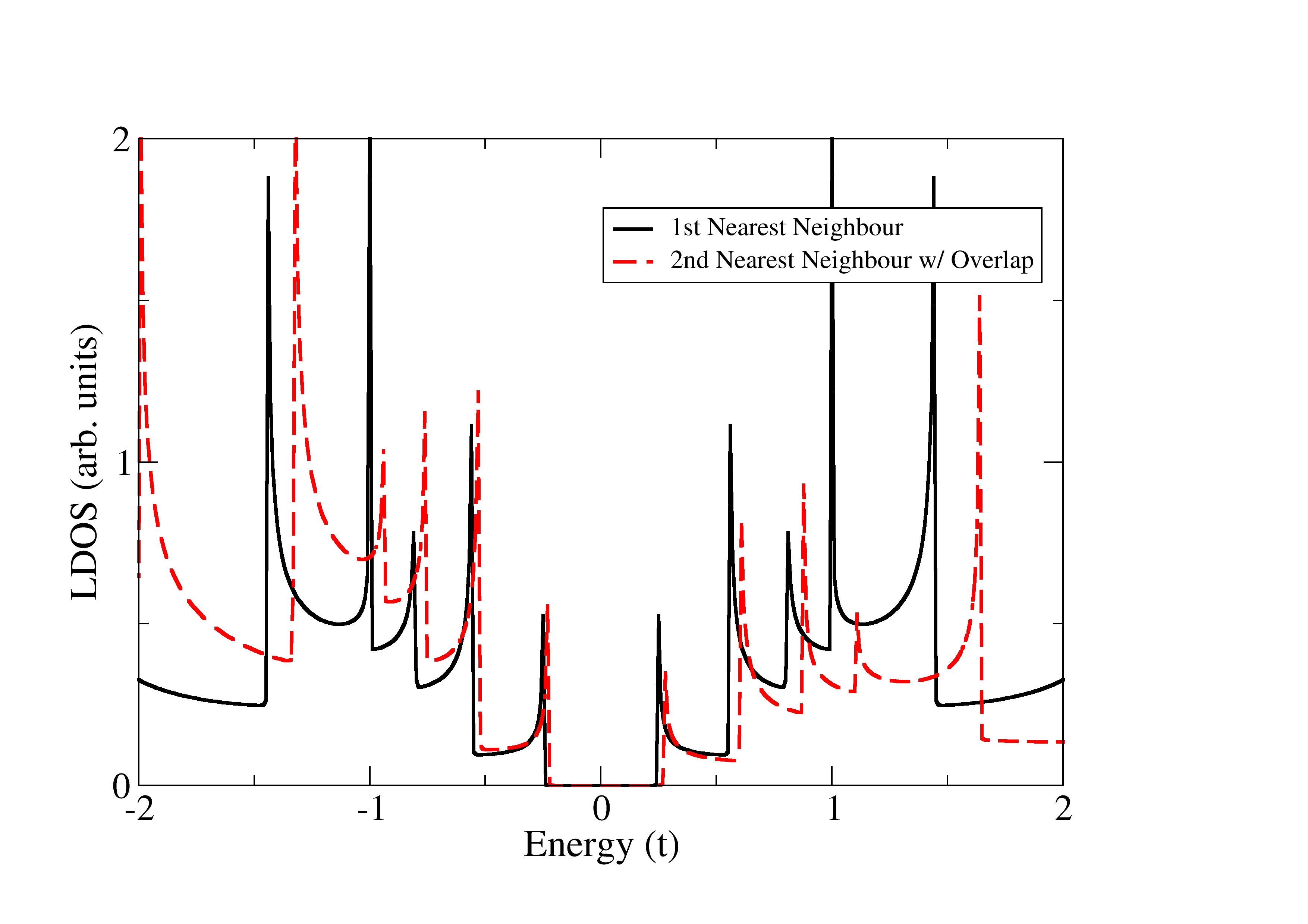}
 \caption{ Local Density of States (LDOS) spectrum of an $N_C = 7$ zig-zag CNT, comparison between 1st Nearest Neighbour and Second Nearest Neighbour models. 
 Generally, the features of the spectrum are quite different between models, yet the behaviour around the band gap remains very similar.}
\label{fig:cnt_ldos}
\end{figure}

 It is useful to compare the band gaps obtained using this model to those using previous methods. 
 Table \ref{tab:1} shows such a comparison, where the band gaps of various semi-conducting zig-zag CNTs have been calculated through
 the conventional 1st Nearest Neighbour approximation, the 2nd Nearest Neighbour with overlap approximation discussed in this paper and a recursive Green's Functions method
 following a similar scheme employed in literature \cite{recursive1, recursive2}. 
 These values are also compared to the most recent accepted band gap values found through the popular B3LYP DFT method \cite{b3lyp}.
 Ultimately, it can be seen that the band gaps calculated using the second nearest neighbour model are only slightly better than those calculated using the first nearest neighbour method.
 Although the behaviour of the LDOS spectrum is drastically different between the two models (see Fig. \ref{fig:cnt_ldos}), the behaviour around the band gap is largely identical. 
 
\section{Stationary Phase Approximation of the off-diagonal lattice Green's Functions}

The Stationary Phase Approximation (SPA) is a method to approximate integrals of the form

\begin{equation}
\label{eq:spa}
  \int dy F(y) e^{i \phi(y) x} \approx \sum_{y_0} F(y_0) e^{i \phi(y_0) x} \sqrt{\frac{2 i \pi }{\phi''(y_0) x}}.
\end{equation}

Here, $y_0$ denotes stationary points in the phase which occur when the condition $\frac{d \phi}{dy} |_{y_0} = 0$ is satisfied. 
Comparing this to $g_{jl}$ as given by Eq. \ref{eq:semi_analytic_gf} it is apparent that the SPA can be used to approximate the off-diagonal Green's Functions
when the sites $j$ and $l$ are separated by a large distance, causing the integrand to oscillate rapidly as $m+n >> 1$.
We demonstrate this solution for the case where $j$ and $l$ are assumed to be separated sufficiently in the armchair direction (where $m=n$, see Fig. \ref{fig:unitcell}),
but indeed this methodology can be applied to other directions.
The analogue between the single integral version of $g_{jl}$ in Eq. \ref{eq:semi_analytic_gf} and the SPA expression in Eq. \ref{eq:spa} becomes more apparent when 
$D = m+n$ which yields

\[  g_{jl} = \frac{i}{2 \pi} \int_{-\pi/2}^{\pi/2} d k_Z \sum_q \sum_\pm  \frac{\gamma_\pm \left|f\right|  e^{i q D }}{\cos k_z \sin q (2 t \left|f\right| \pm (t - Es))} \]
 Further identifying $q_\pm$ as the phase one can identify the stationary points algebraically using the condition $\frac{dq_\pm}{dk_z}|_{q_0} = 0$,
 and the second differential can be found numerically using $\frac{d^2 q_\pm}{d k_z^2} = \lim_{h \to 0} \frac{ q'(k_z + h) - q'(k_z)}{h}$.
 These components form the SPA approximation of the lattice Green Functions, allowing one to calculate $g_{jl}$ with a high degree of accuracy without integration.

\begin{figure}
\centering
 \includegraphics[scale=0.4]{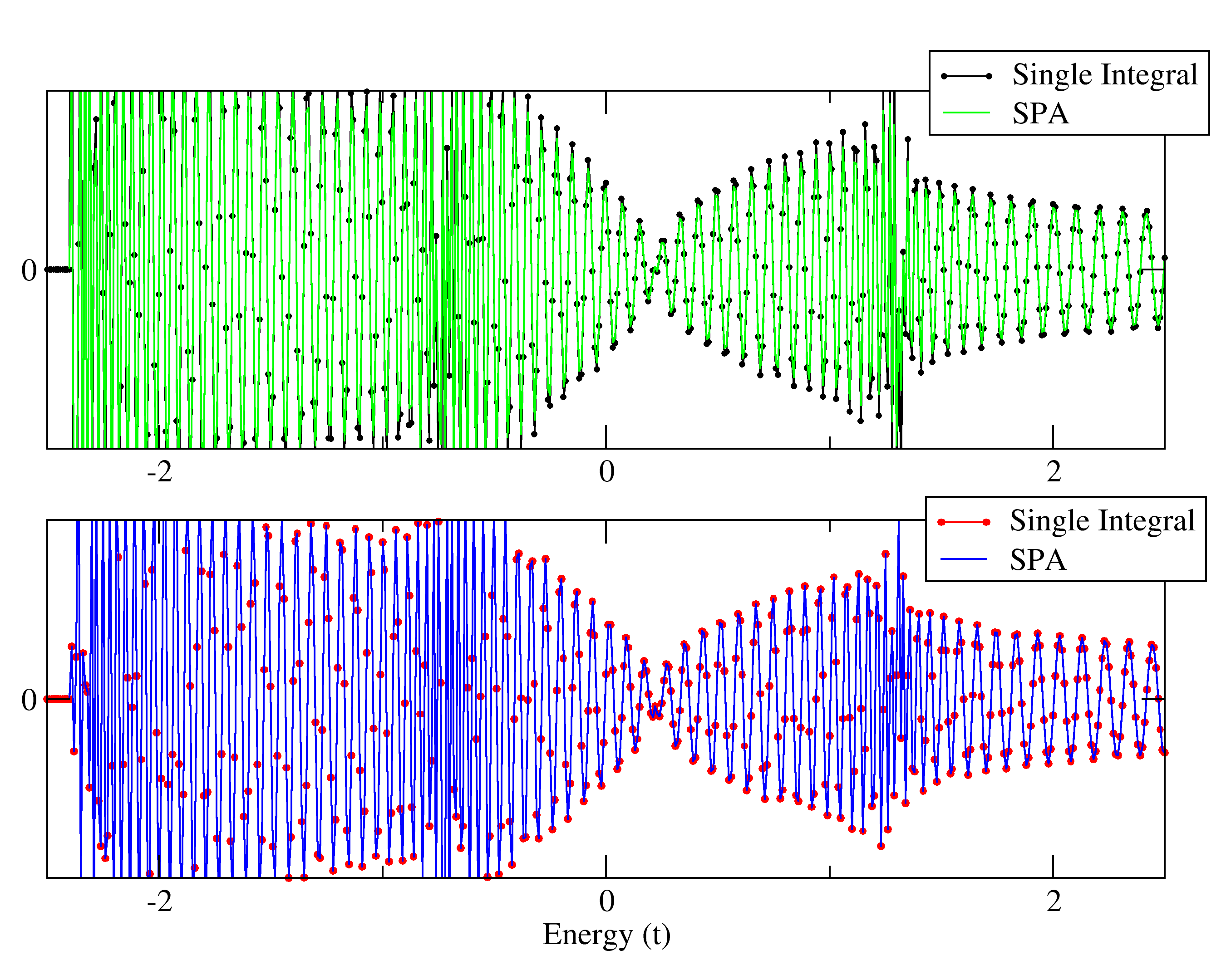}
 \caption{ Comparison of single integral (points) and SPA calculations (continuous) of the Real (top) and Imaginary (bottom) parts of the Green's Function $g_{jl}$ for a separation of $m + n = 80$ in the armchair direction, same sublattice.}
 \label{fig:spa_and_semi_analytic.png}
\end{figure}

 A comparison between the single integral and SPA solution for $g_{jl}$ is demonstrated in Fig. \ref{fig:spa_and_semi_analytic.png} 
 where sites $j$ and $l$ share the same sublattice and are separated by a distance of $D = 80$. 
 This approximation works for all sublattice arrangements and improves in accuracy with increasing separations.

 \section{Conclusions}
 
 In summary, the Green Functions for the graphene lattice accounting for first and second nearest neighbour interactions have been derived 
 with the inclusion of a non-orthogonal basis for the electron wavefunctions, and this was furthermore applied to find a similar expression for the Green's Function of carbon nanotubes.
 We showed how it is possible to simplify the resulting Green Functions from a double to a single integral, which is invaluable for computational purposes.
 In certain cases it is possible to simplify even further, and we showed how a fully analytic expression can be obtained 
 for Green Functions between two distant lattice sites which closely approximates the exact solution.
 The advantage of our extended model is clear when considering extended defects and grain boundaries, which are often dealt with using recursive 
 Green Functions methods in nanoribbons, as such our formalism would allow a more general study of their qualitative behaviour.
 Also, a comparison of our improved model to the rudimentary first nearest neighbour models
 shows that although the resulting behaviour of the GFs around the Dirac point is
 very similar in both cases, the extended model discussed in this paper will be necessary for investigating phenomen at energies beyond the linear dispersion regime.

\section{Acknowledgements}
We acknowledge financial support from the Programme for Research in Third Level
Institutions (PRTLI). MSF also acknowledges financial support from Science Foundation
Ireland (Grant No. SFI 11/RFP.1/MTR/3083).

\end{document}